# Pressure-Induced Phase Transitions in Bilayer La$_3$Ni$_2$O$_7$

Mingyu Xu[1#], Greeshma C. Jose [2#], Aya Rutherford[3], Haozhe Wang[1], Stephen Zhang[4], Robert J. Cava[4], Haidong Zhou[3], Wenli Bi[2*], Weiwei Xie[1*]

[1]Department of Chemistry, Michigan State University, East Lansing, MI, 48824, USA
[2]Department of Physics, University of Alabama at Birmingham, Birmingham, AL, 35294, USA
[3]Department of Physics and Astronomy, University of Tennessee, Knoxville, TN, 37996, USA
[4]Department of Chemistry, Princeton University, Princeton, NJ, 08540, USA

#contributed equally

*Abstract*

La$_3$Ni$_2$O$_7$ exists in two polymorphs: an unconventional structure with alternating layers of single- and triple-layered nickel-oxygen octahedra, and a classical double-layered Ruddlesden-Popper phase. In this study, we report the growth of single crystals of classical double-layered La$_3$Ni$_2$O$_7$ using the floating zone method. Structural characterization under pressures up to 15.4 GPa reveals a gradual transition from orthorhombic to tetragonal symmetry near 12 GPa. Additionally, we present pressure and field-dependent electrical resistance measurements under pressures as high as 27.4 GPa, from which we construct a phase diagram.

**Introduction**

Following the report that $La_3Ni_2O_{7-\delta}$ is superconducting below 80 K at applied pressures of 14–43.5 GPa, significant attention has been focused on nickelate superconductivity. Superconductivity has long been a central focus of scientific investigation, particularly since the discovery of high-temperature cuprate superconductors in the 1980s [1]. The subsequent discovery of Fe-based superconductors [2,3], further expanded the field by demonstrating that the presence of usually magnetic ions like $Fe^{2+}$ does not preclude superconductivity, suggesting a connection between magnetism and superconductivity at high temperatures. Recently, nickelates have garnered substantial attention due to their structural similarities to cuprates and their inclusion of Ni ($3d^8$ in its 2+ state), which is adjacent to Cu ($3d^9$ in its 2+ state) in the periodic table [4]. Significant effort has thus been dedicated to exploring Ni-based superconductors [5–16]. The so-called infinite-layer nickelates, which feature $Ni^+$ ($3d^9$), have been thoroughly studied [8,9,11,12,17]. Early studies identified superconductivity in nickelate thin films, such as Sr-doped $NdNiO_2$, with superconducting transition temperatures ($T_c$) between 6 and 15 K [8,15,18]. More recently, the report of superconductivity in $La_3Ni_2O_{7-\delta}$, with a $T_c$ of 80 K under applied pressures ranging from 14 to 43.5 GPa [19], has been seen by many as a major event, as 80 K exceeds the boiling point of liquid nitrogen and since the compound involves normally magnetic Ni ions, it appears to satisfy many criteria for the identification of new high-temperature superconducting systems. A wave of research activity followed this report [20–26] although limited magnetic shielding remains challenges in the characterization of this system.

At ambient pressure, $La_3Ni_2O_7$ exists in two distinct polymorphs: one composed of alternating single- and triple-layered nickel-oxygen octahedral layers (here referred to as the 1313 phase, space group *Cmmm*), and the other featuring double-layered nickel-oxygen octahedral layers in a classical Ruddlesden Popper phase, (here referred to as the 2222 phase, space group *Cmcm*) [27,28]. While $La_3Ni_2O_7$ in its 1313 form has been proposed as the potential host phase for superconductivity under high pressure [20], the high-pressure behavior of the 2222 phase remains largely unexplored.

In this work, we synthesized $La_3Ni_2O_7$-2222 single crystals using the floating zone method. The single crystals were characterized through single-crystal X-ray diffraction and electrical resistance measurements at various pressures. A pressure-temperature phase diagram is constructed based on these measurements, revealing a structural transition at around 12 GPa, and on cooling this high-pressure phase, a transition near 80 K is seen, which has been attributed to superconductivity by others.

## Experimental

**Single Crystal Growth of $La_3Ni_2O_7$:** Single crystals of $La_3Ni_2O_7$-2222 were grown using the floating zone method in a vertical optical-image furnace. Stoichiometric mixtures of $La_2O_3$ (pretreated at 1000 °C) and NiO were thoroughly ground and fired at 1050 °C for 24 hours. The precursor powders were then hydrostatically pressed into rods and sintered at 1400 °C for 12-24 hours. Crystal growth was performed directly from the sintered rods under a 100% $O_2$ atmosphere at approximately 14-15 bar pressure. During growth, the traveling rate of the crystal boule was maintained at 3-4 mm/h, with the feed rod and seed counter-rotated at 15-20 rpm. $La_3Ni_2O_7$-2222 crystals were successfully extracted from the resulting boule.

**Crystal Structure Determination:** A single crystal with dimensions 0.069 × 0.045 × 0.014 mm³ was selected for analysis. The crystal was mounted on a nylon loop using Paratone oil, enabling examination via a Rigaku XtalLAB Synergy, Dualflex, Hypix single-crystal X-ray diffractometer. Measurements were conducted at room temperature and ambient pressure. Crystallographic data were acquired using the ω scan method with Mo $K_α$ radiation (λ = 0.71073 Å) generated by a micro-focus sealed X-ray tube operating at 50 kV and 1 mA. Experimental parameters, including the total number of runs and images, were optimized algorithmically through strategy computations within the CrysAlisPro software, version 1.171.42.101a (Rigaku OD, 2023). Data reduction incorporated Lorentz and polarization corrections, followed by advanced numerical absorption corrections utilizing Gaussian integration across a multifaceted crystal model [29]. Additionally, empirical absorption corrections using spherical harmonics were applied through the SCALE3 ABSPACK scaling algorithm to further perfect the dataset [30]. The structure was solved and refined using the Bruker SHELXTL Software Package in space group *Cmcm* with Z = 4 [31,32].

**High-Pressure Single-Crystal X-ray Diffraction:** A high-pressure X-ray diffraction (XRD)

experiment was conducted on the same single crystal specimen of $La_3Ni_2O_7$-2222 at applied pressures up to 15.4 GPa. The crystal was first examined at ambient pressure. For the high-pressure measurements, the sample was loaded into a Diacell One20DAC (Diamond Anvil Cell) manufactured by Almax-easyLab, equipped with 500 μm culet-size Boehler-Almax type anvils. A 250 μm thick stainless-steel gasket was pre-indented to 48 μm, and a 210 μm hole was drilled using an electric discharge machining system to accommodate the sample. A 4:1 methanol-ethanol mixture was employed as the pressure-transmitting medium to ensure hydrostatic conditions during the experiment [33]. The pressure inside the cell was monitored using the $R_1$ fluorescence line of a ruby pressure as a calibrant, ensuring accurate pressure measurements throughout the experiment [34–36].

**High-pressure Electrical Resistance Measurements:** High-pressure electrical resistance measurements of $La_3Ni_2O_7$ were conducted in the Quantum Design Physical Measurements System (PPMS-DynaCool). The Electrical Transport Option (ETO) option was used for four-probe electrical resistance measurements. High pressures were achieved using a pair of diamond anvils of 500 μm diameter culet size in a diamond anvil cell made of Be-Cu alloy. Two experimental runs were conducted using AC currents of 0.5 mA at a frequency of 18.3 Hz on two single crystals. A stainless steel gasket, initially 250 μm thick, was pre-compressed to 75 μm, and a 176 μm diameter hole was drilled in its center. To insulate the electrode leads from the metallic gasket, the gasket was coated with a fine mixture of cubic boron nitride powder and epoxy. Slim platinum foils, shaped into thin wedges, were used as electrodes. The $La_3Ni_2O_7$ crystal was placed in center of the gasket hole. Pressures were applied at room temperature and measured through ruby fluorescence method [37].

## Results and Discussion

**Fig. 1a** illustrates the crystal structures of $La_3Ni_2O_7$-2222 at both ambient pressure and 15.4 GPa. With the exception of a change from orthorhombic to tetragonal symmetry, the basic crystal structure is maintained in this pressure range. [Detailed information regarding the ambient pressure X-ray diffraction measurements is provided in the supplementary information (**Tables SI** and **SII**)]. $La_3Ni_2O_7$-2222 has a classical Ruddlesden-Popper bilayer stacking, featuring a pseudo-F-centered orthorhombic lattice with two perovskite-like [i.e. $LaNiO_3$] layers separated by rock salt [LaO] layers [28]. As depicted in **Fig. 1c**, the lattice parameters and unit cell volume decrease with increasing pressure and, as shown in **Figs. 1a** and **1d** at ambient pressure, the Ni-O1-Ni bond angle is approximately 167.8°. As pressure increases to about 12 GPa, this bond angle gradually approaches 180°, while the ratio of the $c$ to $b$ lattice parameters ($c/b$) approaches 1. This suggests a smooth structural transition from an orthorhombic to a tetragonal phase in a second-order transition (within the current measurement resolution), as indicated by the continuous change in the lattice parameters without any abrupt shifts (**Fig 1c**). The volume change as a function of pressure is shown in **Fig. 1e** with the Birch-Murnaghan fitting, $B_0 = 143.6$ GPa.

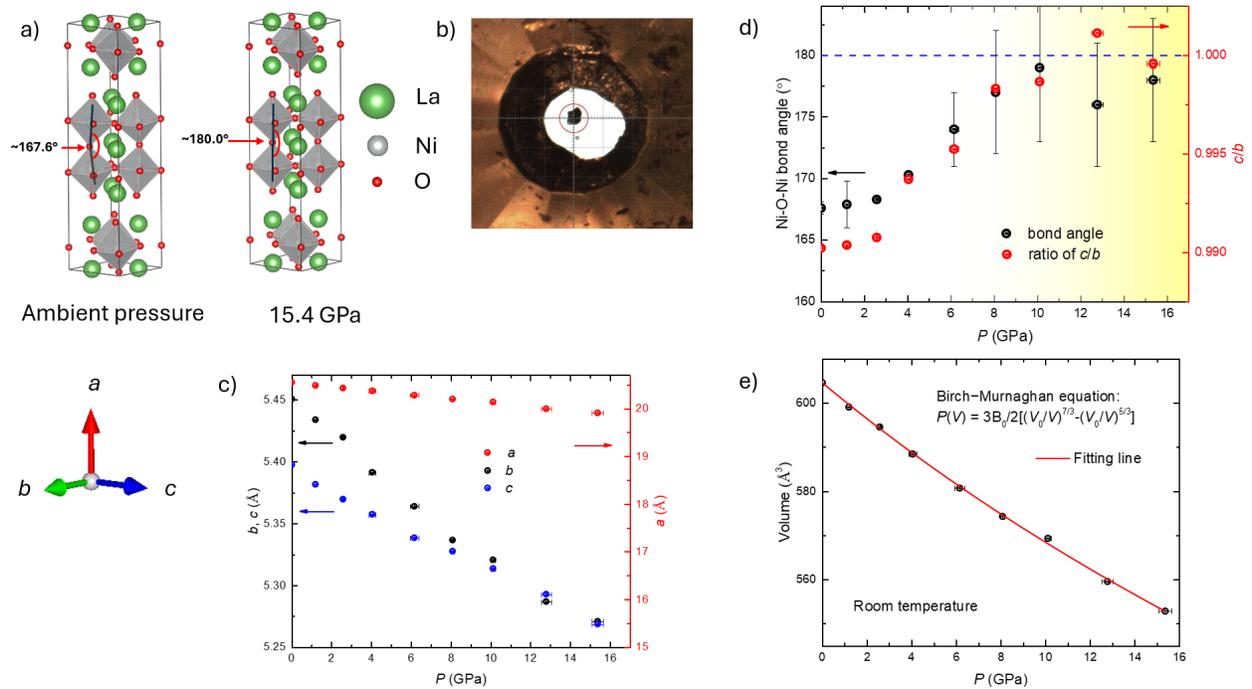

**Fig. 1. Crystal Structure, Lattice Parameter, and Bond Angle Information for La$_3$Ni$_2$O$_7$-2222 under Pressure up to 15.4 GPa.** (*a*) Crystal structure of La$_3$Ni$_2$O$_7$-2222, at 0 GPa and 15.4 GPa. (Visualized using VESTA software [38].) (*b*) The sample was studied inside the Diamond Anvil Cell (DAC) at a pressure of 15.4 GPa. (*c*) The lattice parameters (*a*, *b*, and *c*) as a function of pressure. (*d*) The Ni-O1-Ni bond angle and the ratio of the *c* to *b* lattice parameters (*c/b*) under pressure. The blue dashed line marks the Ni-O1-Ni bond angle of 180° and a *c/b* ratio of 1. The yellow-shaded region highlights the pressure range where a symmetry change has occurred. (*e*) Volume change as a function of pressure with Birch-Murnaghan fitting.

**Figs. 2*a*** and **2*b*** display the temperature dependence of the electrical resistance of a La$_3$Ni$_2$O$_7$-2222 single crystal in different pressure regimes. In the low-pressure range, a kink-like feature appears on cooling around 130 K, accompanied by a broad hump near 220 K, consistent with measurements reported by other groups [25]. The kink-like features have been attributed to spin density wave transitions (T$_{SDW}$) [25,39–42] by others. As the pressure rises, this feature becomes progressively weaker and is eventually suppressed. In the pressure range below 12 GPa, electrical resistance as a function of temperature exhibits insulating behavior over the entire temperature range studied. However, as pressure increases further, the resistance-temperature relationship shifts toward a more metallic character. By 19.3 GPa, the metallic behavior of the resistance becomes evident. At pressures above 12.0 GPa, a pressure where the symmetry of La$_3$Ni$_2$O$_7$-2222 is tetragonal, another, lower temperature feature first

emerges, becomes more pronounced, and then stabilizes near 80 K. This looks like the feature that has been attributed to superconductivity by others, although unfortunately, the resistance does not drop to 0.

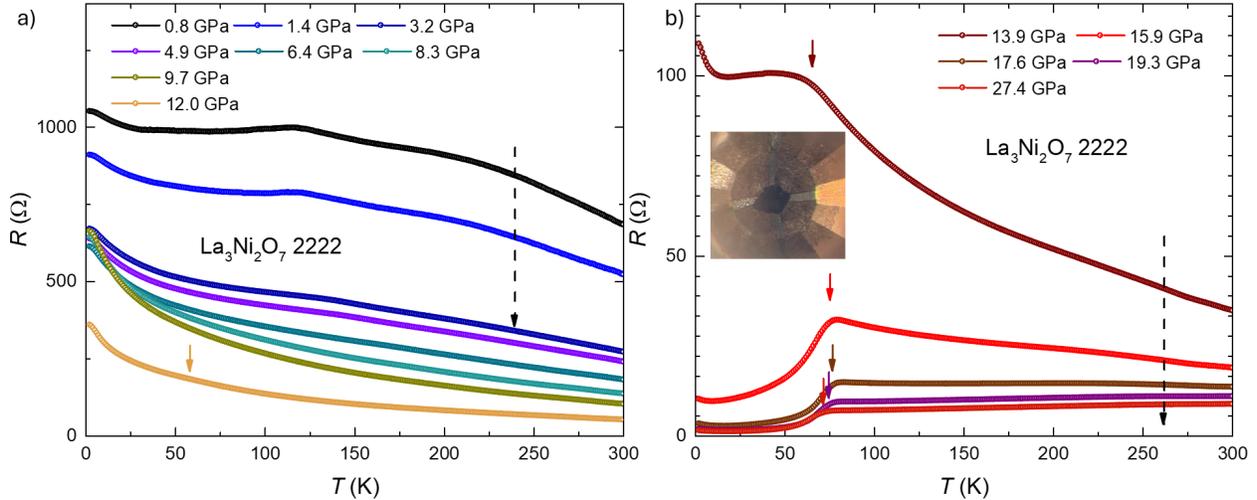

**Fig. 2. Electrical transport measurements of a $La_3Ni_2O_7$-2222 single crystal at pressures up to 27.4 GPa. (*a*).** The electrical resistance of $La_3Ni_2O_7$-2222 as a function of temperature (from 2 K to 300 K) at pressures up to 12.0 GPa. The black dashed line indicates the direction of increasing pressure, while the colored arrows mark the transition temperatures. (*b*). The electrical resistance versus temperature in the pressure range of 13.9 to 27.4 GPa. Inset shows a picture of the sample in the DAC.

**Fig. 3*a*** presents the pressure-temperature (P-T) phase diagram of $La_3Ni_2O_7$-2222, based on the criteria outlined in the supplementary information (**Fig. S1*a*** and **S1*b***). The transition temperatures for what has been identified as a spin density wave transition ($T_{SDW}$) and the new low-temperature transition (T') were determined resistively from the mean values of the onset and transition completion temperatures. The error bars represent the transition widths defined in **Fig. S1*a*** and **S1*b***. $T_{SDW}$ increases with increasing pressure, consistent with recent muon-spin rotation/relaxation (μSR) studies [41]. The transition temperatures observed in both types of experiments are comparable, with minor differences likely attributed to variations in measurement criteria and sample heterogeneity. The spin density wave vanishes at pressures near 10 GPa - the pressure at which the Ni-O1-Ni bond angle reaches 180°, marking a structural transition from orthorhombic to tetragonal symmetry. A new transition (here called T') emerges at approximately the same pressure. T' stabilizes at approximately 80

K after 15.9 GPa and remains nearly constant as the pressure increases up to 27.4 GPa, the limit of our measurements. Finally, **Fig. 3b** shows the electrical resistance at 300 K as a function of pressure, with R (300 K) decreasing significantly as the pressure increases.

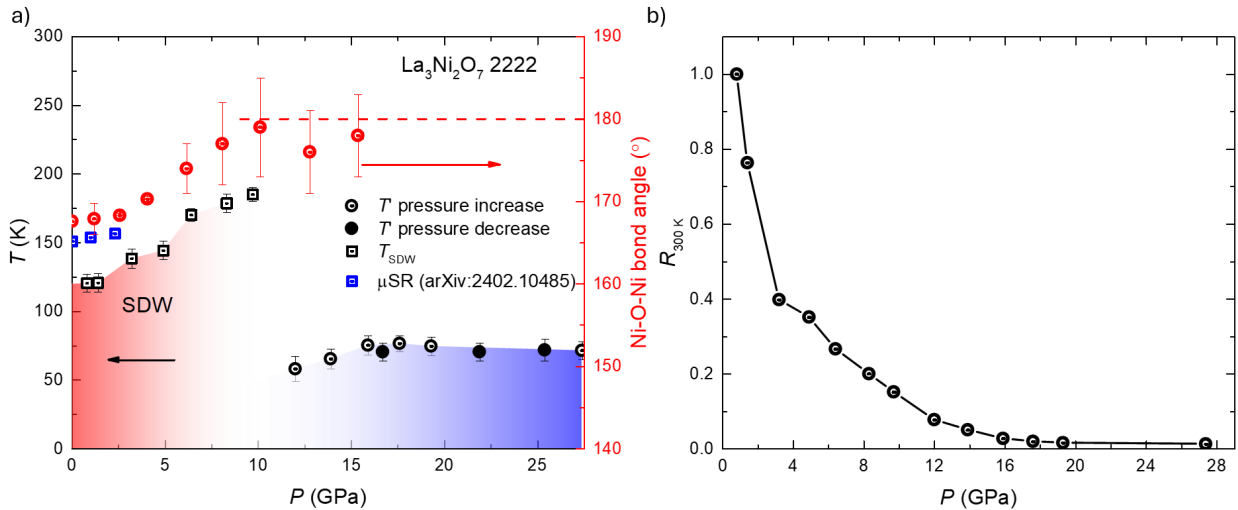

**Fig. 3. Pressure-temperature (P-T) phase diagram and the resistance of a single crystal of 2222-La$_3$Ni$_2$O$_7$ at 300 K as a function of pressure.** (*a*). The P-T phase diagram for La$_3$Ni$_2$O$_7$ -2222, based on our electrical resistance and structural characterization. The black squares represent the spin density wave transition temperature (T$_{SDW}$) determined by high-pressure resistance measurements, while the blue square gives the results of µSR studies [41]. On the right-hand side of **Fig. 3a**, the Ni-O1-Ni bond angles are plotted using red symbols, with the red dashed line indicating the 180° bond angle. The black hollow circles indicate the new transition temperature (T'), while the black solid circles show T' during pressure release **(b).** The electrical resistance at 300 K as a function of pressure for La$_3$Ni$_2$O$_7$-2222.

**Fig. 4a** illustrates the temperature dependence at 27.4 GPa of the normalized resistance (*R*(T)/*R*(150 K)) of a La$_3$Ni$_2$O$_7$-2222 single crystal in applied magnetic fields between 0 and 90 kOe. The sudden drop in resistance at near 80 K has been attributed to superconductivity by others studying La$_3$Ni$_2$O$_7$. The inset emphasizes the fact that the transition temperature does not change much with the applied field. **Fig. 4b** shows the corresponding temperature-field behavior. Both figures demonstrate that the onset values, even using different criteria, of the transition shift slightly with the applied magnetic field. There is thus no clear evidence of a magnetic field suppression of the transition temperature that would normally be observed for superconductivity, especially if it occurs in a very small fraction of the sample, implied by the absence of zero resistance, even though the possibility of filamentary superconductivity still exists. Although the magnetic field does not significantly influence the transition temperature or the resistance prior to the transition, a noticeable change in resistance occurs under the applied magnetic fields, below the onset, the transition broadens. This also occurs in some antiferromagnetic systems such as $Nd_3TiSb_5$ [43]. Similar temperature-dependent resistance behavior under different magnetic fields was observed at 15.9 GPa (**Fig. S2**), further supporting these findings. Together, these results suggest that a magnetic transition, potentially due to the elimination of the magnetic scattering from a spin density wave, exists at around 80 K for $La_3Ni_2O_7$-2222 in the high-pressure range (above 12 GPa), or a filamentary superconducting transition.

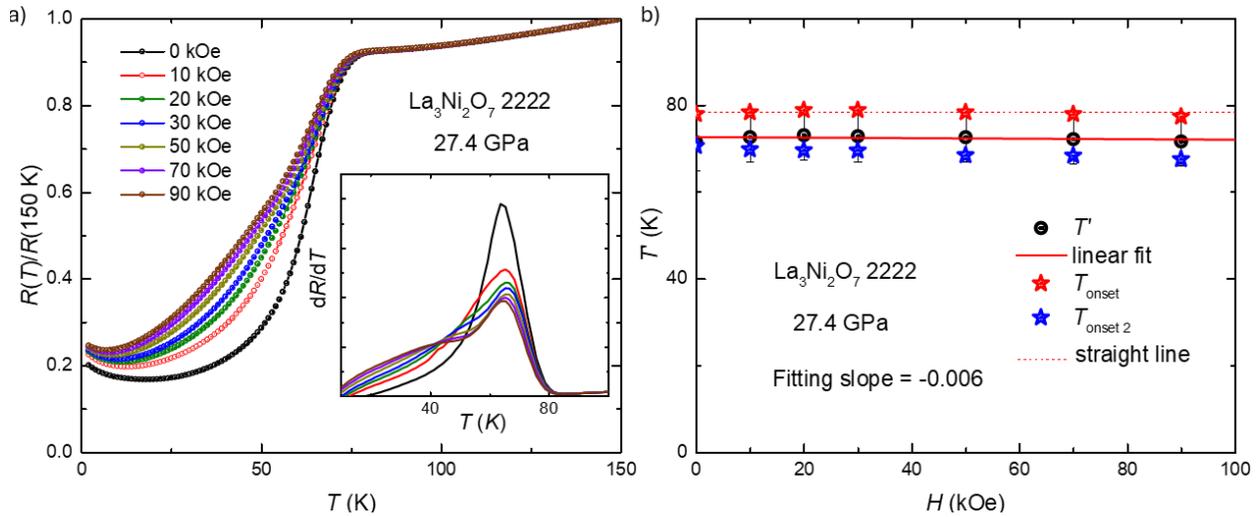

**Figure 4. The normalized resistance under different magnetic fields.** (a) Detail of the normalized resistance ($R(T)/R(150$ K$)$) of $La_3Ni_2O_7$-2222 under different magnetic fields and (b) the $H$-$T$ behavior of the 80 K transition in $La_3Ni_2O_7$-2222 at 27.4 GPa. The black symbols show the transition temperatures determined using the criterion in **Fig. S2**, the $T_{onset}$ is the onset value. $T_{onset\ 2}$ is given by a different criterion, which is 90% of $R$ (77.1 K), which is defined as the highest value before the sudden drop.

## Conclusion

$La_3Ni_2O_7$-2222 single crystals were grown using the floating zone method and subjected to high-pressure single-crystal X-ray diffraction and electrical resistance measurements. These experiments allowed for the construction of a temperature-pressure phase diagram. The ambient pressure phase is stable up to about 12 GPa, consistent with previous μSR results [41]. At about 12 GPa, a structural transition from orthorhombic to tetragonal symmetry is observed in $La_3Ni_2O_7$-2222, accompanied by an 80 K phase transition on cooling. This transition does not exhibit characteristics typical of superconductivity, as the transition temperature remains largely unaffected by an applied magnetic field, and the material displays insulating or semiconducting behavior prior to the transition at 15.9 GPa, however, the possibility of filamentary superconductivity still exists. Given the significant resistance drop and the presence of SDW at lower pressures, we speculate that this transition is likely a magnetic transition that reduces the magnetic scattering or a filamentary superconducting transition. We also point out that the temperature of the 80 K transition is only slightly changed in the range of pressures between about 15 and 25 GPa, which, along with the independence of the magnetic field, suggests that strong magnetic forces are involved. Further experiments are clearly required to determine the nature of the high-pressure $La_3Ni_2O_7$-2222 phase below 80 K.

## Acknowledgments

The work at Michigan State University was supported by the U.S.DOE-BES under Contract DE-SC0023648. W.B. and G. C. J. were supported by the National Science Foundation (NSF) CAREER Award No. DMR-2045760. The Physical Properties Measurements System (PPMS) employed in the high-pressure electrical transport experiments was acquired under NSF MRI Grant No. 2215143. The work at the University of Tennessee was supported by the Air Force Office of Scientific Research under grant no. FA9550-23-1- 0502. The work at Princeton University was supported by the US Department of Energy, Division of Basic Energy Sciences, grant DE-FG02-98ER45706.

# Supporting Information

# Pressure-Induced Phase Transitions in Bilayer $La_3Ni_2O_7$

Mingyu Xu[1#], Greeshma C. Jose[2#], Aya Rutherford[3], Haozhe Wang[1], Stephen Zhang[4], Robert J. Cava[4], Haidong Zhou[3], Wenli Bi[2*], Weiwei Xie[1*]

[1]Department of Chemistry, Michigan State University, East Lansing, MI, 48824, USA
[2]Department of Physics, University of Alabama at Birmingham, Birmingham, AL, 35294, USA
[3]Department of Physics and Astronomy, University of Tennessee, Knoxville, TN, 37996, USA
[4]Department of Chemistry, Princeton University, Princeton, NJ, 08540, USA

#contributed equally

**Table SI.** The crystal structure and refinement of La$_3$Ni$_2$O$_7$ at room temperature K (Mo Kα radiation). Values in parentheses are estimated standard deviation from refinement.

| Chemical Formula | La$_3$Ni$_2$O$_7$ |
|---|---|
| Formula Weight | 646.15 g/mol |
| Space Group | *Cmcm* |
| Unit Cell dimensions | $a$ = 20.5510(13) Å<br>$b$ = 5.4508(4) Å<br>$c$ = 5.3975(4) Å |
| Volume | 604.62(7) Å$^3$ |
| Z | 4 |
| Density (calculated) | 7.098 g/cm$^3$ |
| Absorption coefficient | 26.849 mm$^{-1}$ |
| F (000) | 1132 |
| 2θ range | 7.734 to 81.316° |
| Reflections collected | 4856 |
| Independent reflections | 1047 [R$_{int}$ = 0.0701] |
| Refinement method | Full-matrix least-squares on F$^2$ |
| Data/restraints/parameters | 1047/0/37 |
| Final *R* indices | $R_1$ (I>2σ(I)) = 0.0460; $wR_2$ (I > 2 σ(I)) = 0.1067<br>$R_1$ (all) = 0.0862; $wR_2$ (all) = 0.1261 |
| Largest diff. peak and hole | +7.82 e$^-$/Å$^3$ and -2.74 e$^-$/Å$^3$ |
| R. M. S. deviation from mean | 0.769 e$^-$/Å$^3$ |
| Goodness-of-fit on F$^2$ | 0.997 |

**Table SII.** Atomic coordinates and equivalent isotropic atomic displacement parameters (Å$^2$) of La$_3$Ni$_2$O$_7$. ($U_{eq}$ is defined as one-third of the trace of the orthogonalized $U_{ij}$ tensor.) Values in parentheses are estimated standard deviations from refinement.

| Atoms | Wyck. | x | y | z | Occ. | U$_{eq}$ |
|---|---|---|---|---|---|---|
| La1 | 4c | 0 | 0.75046(11) | 1/4 | 1 | 0.008(14) |
| La2 | 8g | 0.32020(2) | 0.25758(7) | 1/4 | 1 | 0.007(13) |
| Ni | 8g | 0.09594(5) | 0.25201(16) | 1/4 | 1 | 0.005(19) |
| O1 | 4c | 0 | 0.29120(15) | 1/4 | 1 | 0.008(15) |
| O2 | 8g | 0.20360(3) | 0.21760(12) | 1/4 | 1 | 0.012(13) |
| O3 | 8e | 0.39500(4) | 0 | 0 | 1 | 0.013(13) |
| O4 | 8e | 0.08990(3) | 0 | 0 | 1 | 0.012(14) |

**Tables SI and SII** show the results of the single-crystal XRD. The structure was solved and refined using the Bruker SHELXTL Software Package with the space group *Cmcm*, $La_3Ni_2O_7$. The final anisotropic full-matrix least-squares refinement on $F^2$ with 37 variables converged at $R_1$ = 8.62%, for the observed data and $wR_2$ = 12.61% for all data. The goodness-of-fit was 0.997. The largest peak in the final difference electron density synthesis was 7.82 e⁻/Å³, and the largest hole was -2.74 e⁻/Å³ with an RMS deviation of 0.769 e⁻/Å³. Based on the final model, the calculated density was 7.098 g/cm³ and F (000), 1132 e⁻.

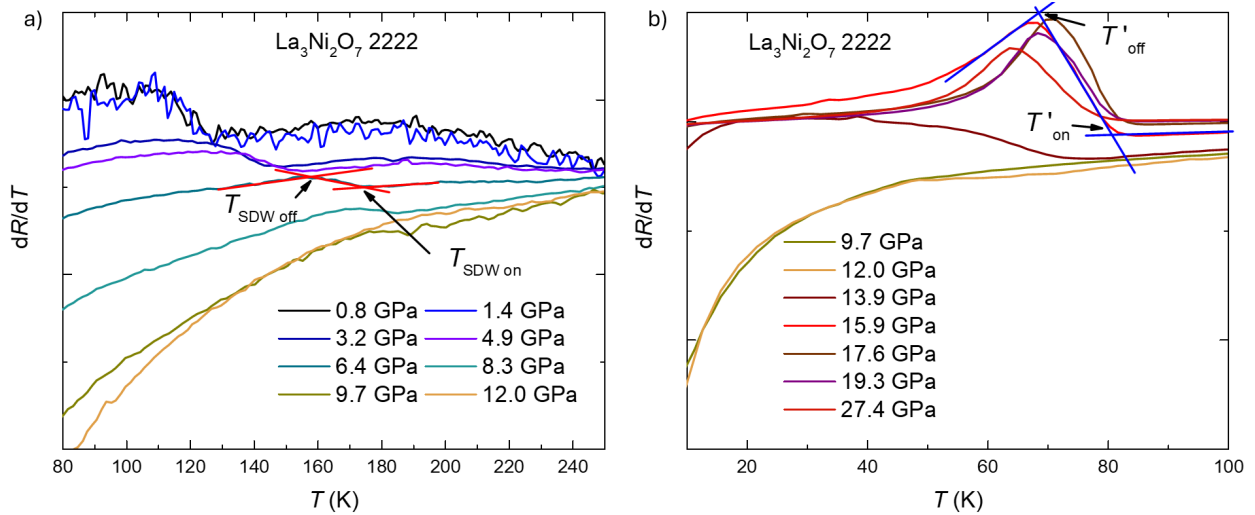

**Fig. S1. Criterion of T$_{DW}$ and T′ based on the plots of d$R$/d$T$ at different pressures. Fig. S1a** shows the criterion (onset and offset value) of T$_{SDW}$ using d$R$/d$T$ in the temperature range from 80 K to 150 K. **Fig. S1b** gives the criterion of T′ based on d$R$/d$T$ in the temperature range from 10 K to 100 K.

Fig. S1 shows the criteria for the determination of $T_{SDW}$ and $T'$. The values of $T_{SDW}$ and $T'$ are given by the mean values of onset and offset, and the transition width is given by the half value of the difference of onset and offset.

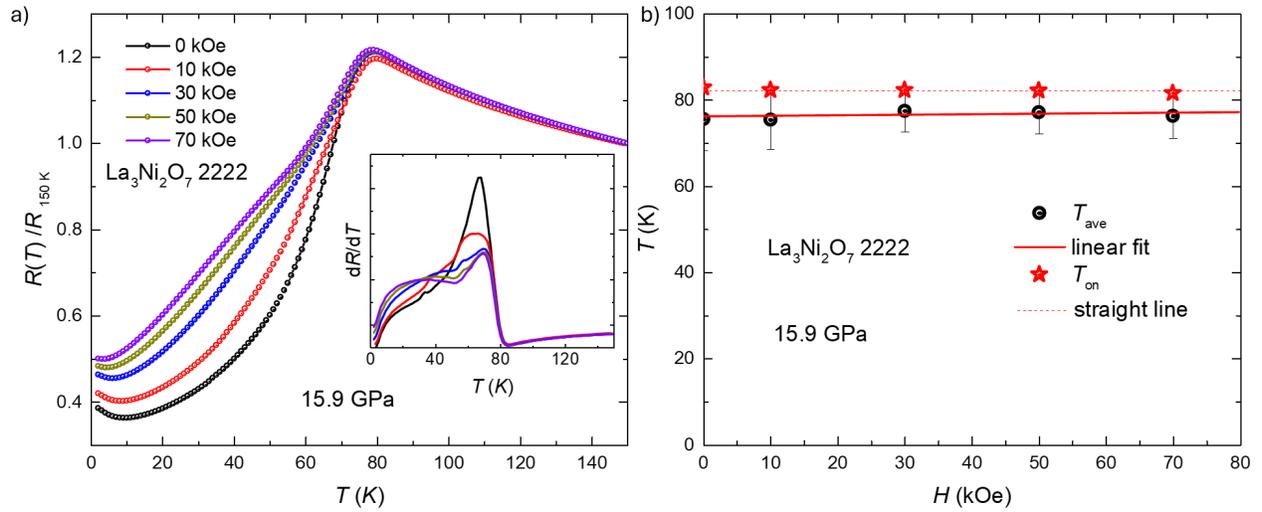

**Fig. S2. Normalized resistance ($R(T)/R(150\ K)$) of La$_3$Ni$_2$O$_7$-2222 under the different magnetic fields (Fig. S2a) and $T'$-$H$ phase diagram (Fig. S2b) at 15.9 GPa.** Fig. S2a shows the normalized resistance as a function of temperature under the different magnetic fields at 15.9 GPa. (Inset) The derivative of resistance as a function of temperature. **Fig. S2b** shows the temperature-magnetic field ($T$-$H$) phase diagram. The black symbols show the $T'$ values, and the red star symbols present the onset values. The red solid line gives the linear fit of $T'$, and the red dashed line presents the straight line across the onset values.